# QoE Modeling for Voice over IP: *Simplified E-model Enhancement Utilizing the Subjective MOS Prediction Model – A Case of G.729 and Thai Users*

Therdpong Daengsi
Dept. of Sustainable Industrial Management Engineering
Faculty of Engineering
Rajamangala University of Technology Phra Nakhon
Bangkok, Thailand
therdpong.d@rmutp.ac.th; +66814965464

Pongpisit Wuttidittachotti
Dept. of Data Communication and Networking
Faculty of Information Technology
King Mongkut's University of Technology North Bangkok
Bangkok, Thailand
Pongpisit.w@it.kmutnb.ac.th; +66956419959

*Abstract*—This research proposes an enhanced measurement method for VoIP quality assessment which provides an improvement to accuracy and reliability. To improve the objective measurement tool called the simplified E-model for the selected codec, G.729, it has been enhanced by utilizing a subjective MOS prediction model based on native Thai users, who use the Thai-tonal language. Then, the different results from the simplified E-model and subjective MOS prediction model were used to create the Bias function, before adding to the simplified E-model. Finally, it has been found that the outputs from the enhanced simplified E-model for the G.729 codec shows better accuracy when compared to the original simplified E-model, specially, after the enhanced model has been evaluated with 4 test sets. The major contribution of this enhancement is that errors are reduced by 58.87 % when compared to the generic simplified E-model. That means the enhanced simplified E-model as proposed in this study can provide improvement beyond the original simplified one significantly.

*VoIP; Simplified E-model; G.729; QoE; MOS;Thai users*

## I. Introduction

Voice over Internet Protocol (VoIP) is the convergence of computer network technology, especially the IP networks at present, and telecom telecommunication. Nevertheless, because of the philosophy of voice communication that is a kind of real-time communication, voice quality may be affected by the impairment factors (e.g. loss and delay) that might occur in IP networks from poor network planning (e.g. an IP network without Quality of Service management) [1]. Therefore, based on Quality of Experience (QoE) modeling for VoIP, this research presents the proposed method to improve reliability and accuracy of an objective method for VoIP quality measurement, called the simplified E-model for G.729 which is a popular codec used in VoIP networks. It is the modified and extended version of [2] with additional data from an extra scenario and participants. The advancement beyond the previous version is the assessment result from the enhanced simplified E-model utilizing Subjective MOS Prediction Model that shows 71 % improvement when compared to the original simplified E-model. Furthermore, this approach can be adapted for users in other countries that people use other languages to improve a generic simplified E-model.

For the structure of this article, there are six sections. After the introduction, background information are presented in Section II, including the G.729 codec, impairment factors, quality of experience, subjective quality measurement, simplified E-model and subjective MOS prediction model. Section III covers related research works. Section IV and V describe development of the subjective MOS prediction model and the simplified E-model enhancement respectively. Section VI describes the performance evaluation of the enhanced model, followed by a discussion in Section VII. Last but not least, the future work and conclusion are described in Section VIII. Section I now continues with a VoIP overview and motivation of works follows:

### A. VoIP Overview

VoIP has become the fastest growing internet application in the past several years [3]. In this era of VoIP technology, there are mainly four forms or methods using VoIP, as follows [4]:

- Computer-to-computer: it is the most effective and easiest method to use VoIP via applications, such as, Skype, Hangout, Viber, and LINE. These require users to have a microphone and speakers. Mobile devices with VoIP applications are classified in this form.
- Telephone-to-telephone: this method is a form of IP-Private Branch Automatic Exchange (IP-PABX) by multiple companies which can save cost when calling among their own branches located in different cities, for example.
- Telephone-to-computer: it is an option that companies provide special numbers or calling cards, in order to allow users to utilize a traditional telephone to make a call to a user on a computer or a mobile device with a VoIP application installed.
- Computer-to-telephone: this method is similar to computer-to-computer. With a VoIP application and services (e.g. Skype), users can make calls from a VoIP application to a user on a traditional telephone with cheap rate service charges.

VoIP communications offer users the possibility to make free (or low cost) calls utilizing IP networks [5]. However, VoIP quality has been highlighted as a dark point of VoIP systems/applications [5]. This is one of reasons to measure or predict it, in order to look back and improve influential factors leading to better VoIP quality provided to end users.

*B. Motivation of Work*

The motivation of this work is a combination of two issues, the voice quality method improvement and the special characteristic of Thai brains that respond to sounds of Thai speech, in order to have a mathematical model that can be applied to implement a VoIP quality measurement tool for VoIP quality monitoring within Thai environments. Further information is described as follows:

1) To predict VoIP quality using a simple method/tool, the simplified E-model, which has been derived from the original E-model, is an interesting approach. Namely, this model can be utilized to predict VoIP quality by using few factors for calculation [6-7], instead of using many factors like the original E-model. To predict VoIP quality for transmission planning, in term of non-intrusive approach, the E-model is well-known for VoIP quality assessment because it does not require reference signals, degraded signals and signal processing like the Perceptual Evaluation of Speech Quality (PESQ), which is well-known in terms of intrusive method [1, 8]. The E-model was developed by the European Telecommunications Standards Institute (ETSI) in its first phase, before being promoted as a standard by the Telecommunications Standardization Sector of ITU, or called ITU-T for short [9-10]. Nonetheless, the E-model is not a simplex model because it is a calculational model referring to many parameters (e.g. random loss probability and room noise at each side). Therefore, to utilize the generic E-model with simplicity, the simplified E-model has been proposed to use only the factors associated with codec, loss and delay, to calculate the output. Like the original E-model that requires enhancement to gain reliability and high accuracy still because it requires verification from field surveys or laboratory tests for the large number of possible and reasonable conditions that may occur in IP networks [11], therefore, the simplified E-model will also require improvement for better accuracy and reliability.

2) This research is based on a tonal language called the Thai language. It is different from western languages, especially, there are 5 tones, consisting of low-tone, mid-tone, high-tone, rising tone and falling tone [12]. It has been mentioned in [13-14] and proven that native Thai users have better perception to Thai speech than non-Thai native speakers (e.g. Chinese and British) because the left frontal operculum of a Thai brain is activated when listening to Thai speech sounds, which is their mother language, whereas the non-Thai native brain is not activated during the same scenario tests. The main reason is the tonal language which is a key feature of the Thai language can change the meaning of Thai words. For example, "คาง" (mid-tone) means "a shin" but "ค้าง" (high-tone) means "to stay overnight". It is consistent with the evidence as shown in [14] about the comparative results between the ratio of fundamental frequency ($F_0$) and its standard deviation (SD) of Thai speech sounds and the ratios from English sounds.

The combination of these two issues is consistent with the issue that has been acknowledged by ITU-T that subjective quality measurement of multimedia, such as VoIP, can be impacted by the language effects; this issue is now being study by one study group on behalf of ITU-T [15-17]. Therefore, based on the two described issues, the mathematical model became the main focus of this study and to then apply it appropriately in Thailand for a VoIP quality prediction development tool in the future.

## II. BACKGROUND

*A. VoIP Codec: G.729*

As referred in [18] more than 90 % of VoIP traffics, which uses ITU-T codecs, use the G.729 codec while the rest of the VoIP traffic uses G.711. For the G.729, it was issued by ITU-T [19]. Unlike G.711 which is always used for communication within the same area networks, G.729 is a codec commonly utilized over wide area networks (WANs). G.729 uses the coding technique called the Conjugate Structure - Algebraic code-excited linear prediction (CS-ACELP). It requires only 8 Kbps for its payload. One of the major advantages of G.729 codec is requirement of lower bandwidth than G.711 codec because it provides right timing with more compression. This is reason it has been recommended to be utilized over WANs [20]. In the G.729 codec family, G.729A codec is the reduced complexity version of the original G.729. There is built-in packet loss concealment. However, for G.729B codec, it supports Comfort Noise Generation (CNG) function and Voice Activity Detection (VAD) function. It has been mentioned that MOS-LQS of G.729 codec is 4.18, while its MOS-LQO is 3.92 approximately, calculated from the average MOS values from several languages, such as, English, German and Spanish [21-22].

### B. Impairment Factors

VoIP is sensitive easily impacted by impairment factors in the IP network, thus, it requires quality of service guarantees [3]. In IP networks, two major impairment factors are delay and loss that quite impact VoIP quality [13, 23]. In general, loss occurs when packets are transmitted but a few are not received at the receiver due to events occurring in the IP network. The major causes of loss are fiber link damage and router failures in the IP networks [24-25]. Packet loss can be either a random pattern or a bursty pattern, see Fig. 1 [26]. High burst loss impacts voice quality more than low burst loss [26]. The performance of VoIP calls will suffer greatly if loss occurs. For very good VoIP quality, the loss of VoIP packets should be less than 1 %. The loss rate of 3 % or less might be acceptable for the business quality level [1, 27]. Nonetheless, if the loss rate is more than 5% it is considered harmful and a cause of VoIP quality issue.

For packet delay, it is the combination of several kinds of delay including packetization delay, encoding and decoding delay, and network delay. In a few scenarios, packet delay could be increased by conversion of one codec into another codec in a different IP network, and transmission of voice packet through gateways, switches and routers [1]. In general, delays impact VoIP quality at the receiver, particularly a value of 400 ms is recommended and accepted as the maximum delay widely for network planning purposes [28].

### C. Quality of Experience

Quality is an important subjective concept. It is referring to the level of satisfaction of a user to something with an experience. If a user is not satisfied by his/her experience about multimedia services (e.g. text, video and VoIP), for example, the utility of the service dropped [29]. In order to predict and/or measure the quality of those services perceived by end users and to measure performance of systems/networks at the point of view of end users, the new concept called Quality of Experience (QoE) has been proposed instead of Quality of Service (QoS) [30]. Unlike QoS that is mainly related to technical parameters (e.g. delay and loss), QoE focuses on the user's perception and user's expectations towards a presentation or an application because each user normally cares about the experience his/her is able to obtain [31].

QoE is fundamentally a subjective measurement metric. It has become an important key indicator to represent experience of the end user. This provides the quantitative link to user perception because subjective QoE measurement is the most accurate method to obtain the user's perception [32]. Therefore, QoE evaluation or assessment is widely used to measure the level of user's perception or satisfaction provided by multimedia services/applications, which includes VoIP. All system elements involved in the end-to-end service delivery influence the perception of the quality [33]. Besides, psychological, environmental, and sociological factors additionally affect QoE assessment, including expectations of users, user profile and experience. The most popular measure of QoE is mentioned as MOS which stands for the Mean Opinion Score [30]. Particularly, for VoIP quality measurement, MOS has been used and standardized by ITU-T [34-35].

### D. Subjective Quality Measurement and Subjective MOS Prediction Model

Although VoIP quality subjectively requires judgement by users for the most reliability as it is fundamentally determined by the users' perception, there are a few drawbacks (e.g. cost, complexity and time consumption to organize a group of participants or subjects to participate VoIP assessment utilizing subjective tests) [36]. While 'strong variations' and 'trackability problem' may occur during the interaction in each subjective test session, subjective tests are normally considered to be more credible than other assessment methods and used for verification and calibration of those objective measurement methods [37].

For subjective quality measurement methods, listening test results seem more stable and more reliable than other subjective methods because it makes participants concentrate when alone in a special room, called a soundproof room [36]. Each speech sample is voted with an integer between 1 and 5 (5 means excellent whereas 1 means bad) by a number of participants and then MOS is finally computed [1, 32, 36]. Nevertheless, because of reaching the same standard of realism covering delay and loss effects, the subjective test method called conversation opinion test becomes an acceptable option [34-35].

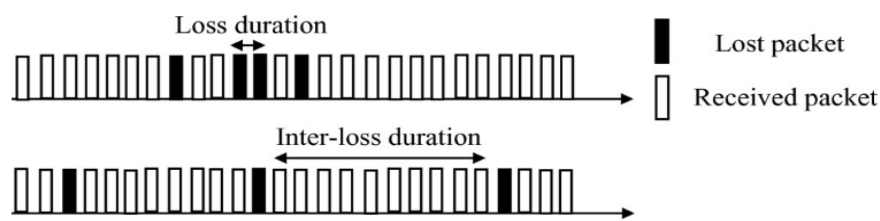

Fig. 1. Examples of random packet (the lower) and bursty packet loss (the upper), adopted from [24]

Previously, the subjective MOS prediction models based on native Thai users referring to delay effects and/or loss effects were proposed in [13, 39, 41]. Its concept is to gather scores from voting by a group of participants or subjects after VoIP quality testing under loss and/re delay scenarios. Then regression techniques are used to derive the mathematic equation [42]. Although it is not an ITU-T standard, it has been mentioned in [13] that the regression based model derived from subjective MOS values from a large group of participants or subjects can provide reliability and accurate improvements when compared to an ITU-T standard model.

### E. Objective Quality Assessment: Simplified E-model

The E-model is widely used because it has been recognized as one of the most popular methods for objective assessment methods [41]. Previously, the E-model is a widely utilized method especially in the prior research directions about speech or voice quality assessment. Between the years 2001 to 2017, there have been 176 published articles associated with the term E-model [44]. This model can be presented as in (1) and (2) [10]. Nevertheless, impairment factors as shown in (1) are not simplex. MOS value, which is a five-point scale, can be converted from the R-value utilizing (2); (Notes: MOS is 4.5 if R is greater than 100, whereas it is 1 if R is less than 0). However, the R-value can be converted from MOS value using (3) [9-11, 14].

$$R = Ro - Is - Id - Ie_{eff} + A \quad (1)$$
$$MOS = 1 + 35R/10^3 + R(R-60)(100-R)7*10^{-6} \quad (2)$$
$$R = 3.026MOS^3 - 25.314MOS^2 + 87.060MOS - 57.336 \quad (3)$$

According to equations above, $R$ stands for R-value, $Ro$ stands for signal-to-noise-ratio, $Is$ stands for simultaneous impairment factor, $Id$ stands for delay impairment factor, $Ie_{eff}$ stands for equipment impairment factor (e.g. packet loss), $A$ stands for advantage factor and $MOS$ stands for Mean Opinion Score.

Nevertheless, because of the complexity of the simplified version of the E-model has been presented in [45]. It is simpler than the original E-model, as shown in (4).

$$R = R_o - I_d(d) - I_e(CODEC, PLR) + A \quad (4)$$

Where $R$ stands for R-value, $Ro$ is 94.2, $I_d$ stands for the delay impairment factors, while $d$ means the average one-way delay in milliseconds. $I_e$ stands for the equipment impairment factor, while CODEC and PLR are the used audio encoding scheme and the packet loss ratio respectively. $A$ stands for the advantage factor, it permits for compensation of impairment factors when the user benefits from other types of access.

However, as recommended in the ITU-T Rec. G.107 [10], the $R_o$ has slightly changed from 94.2 to 93.2, therefore, the other adopted E-model, called the simplified E-model with the updated value of $R_o$, has been considered and presented as in (5), as follows [6-7]:

$$R = Ro - Icodec - Idelay - Ipacketloss \quad (5)$$

Where $Ro = 93.2$ [2, 10-11], $Icodec$ is the codec impairment factor (for G.729, it is 10) [46], $Idelay$, can be applied from (6), and $Ipacketloss$, the packet loss impairment factor, can be applied from (7) respectively, as follows [47]:

$$Idelay = I_d(d) = 24d/10^3 + 11(d-177.3)/10^2 * H(d-177.3) \quad (6)$$
$$Ipacketloss = a + b\ln(1+0.001cP) \quad (7)$$

Where H(x) means the step function where H(x) = 0 if x is less than 0 and 1 otherwise, for $Idelay$ calculation in (6). However, the important caution about using $Idelay$ is the recommendation about using one-way delay about 0-400 ms only [6, 48]. $Ipacketloss$ is the calculated result from (7) [47]. Where P stands for the percent of loss rate (%), a = 10, b = 25.21 and c = 20.20, adopted from [47].

## III. RELATED RESEARCH WORKS

There are about 15 research works focusing on the simplified E-model and the original E-model improvement or enhancement that have been presented previously, including a few previous research works of the authors, as shown in Table 1 [6-7, 14, 43, 47, 49-59]. From the knowledge obtained from previous works, several research works have proposed new factors (e.g. jitter impairment factors) to add into the E-model, while a few of them proposed solutions to improve the simplified E-model. Particularly, it has been shown in [14] that the enhanced simplified E-model for G.711 can provide error reduction of about 46.8 % when compared to the generic simplified one.

Therefore, one can see that there is a gap to modify the generic E-model and its simplified one, especially by utilizing subjective MOS approach associated with packet loss and packet delay for G.729 codec as in this research work, while the combination of subjective MOS prediction model and the simplified E-model for G.729 codec has not been previously studied in depth.

## IV. SUBJECTIVE MOS PREDICTION DEVELOPMENT

To develop the subjective MOS prediction model, subjective test results are commonly required. In this study, the raw data of subjective MOS provided by G.729 have been adopted from one previous work [13]. The validated data was obtained from conversation-opinion tests utilizing Richard's tasks (see Fig 2) with 354 KMUTNB students, referring to 13 scenarios with delay of 0 – 800 ms and loss of 0 - 10 %.

TABLE 1 LIST OF PREVIOUS RESEARCH WORKS REFERRING TO THE SIMPLIFIED E-MODEL AND THE E-MODEL ENHANCEMENT

| References | Proposed concept |
|---|---|
| Ding & Goubran (2003) [49] | $Ie$ and $Ij$ proposed the as shown in (8) and (9) respectively, where $Ie$ stands for the improved $Ie$, $Ie\_opt$ stands for the optimum (without loss) $Ie$, $loss\_rate$ stands for the amount of loss in percent and $C_1$, $C_2$ as in (8) are simultaneous impairment factor. While K stands for a time constant and $C1$, $C2$, $C3$, $C4$ as in (9) are constants.<br>$Ie = Ie\_opt+C_1 \cdot ln(1+C_2 \cdot loss\_rate)$ (8)<br>$Ij = C1 \cdot H^2+C2 \cdot H+C3+C4 \cdot e^{-T/K}$ (9) |
| Zhou et al. (2006) [47] | The prediction of VoIP quality with E-model was proposed. However, after considering in detail, this study actually used a simplified E-model to predict the results because it presented (5)-(6) which are parts of the simplified E-model, while a-c codec constants for G.711, G.729, G.723.1 and iLBC utilized in (6) were also presented. |
| Takahashi et al. (2006) [50] | $MOS\_CQE\_j$ was proposed as shown in (10), where $MOS\_CQE\_j$ stands for a Japanese version of $MOS\_CQE$, whereas $MOS\_CQE$ stands for the generic conversational quality score provided by E-model.<br>$MOS\_CQE\_j=(8681MOS\_CQE+271)/10000$ (10) |
| Ren et al. (2008) [51] | $Il$ was proposed as shown in (12), where $Il$ stands for language impairment factor, $C1, C2$ are constants for Chinese, 0.52819 and 0.574391 respectively and $PPL$ stands for packet loss rate.<br>$Il = C1+C2*PPL$ (11) |
| Ren et al. (2010) [52] | $Ij$ was proposed as in (12), where $\gamma_0$, $\gamma_1$, $\gamma_2$, $\gamma_3$, $\gamma_4$, $\gamma_5$ are constants and $\gamma_0$ stands for the threshold, and H(x) stands for the step function: H(x) = 0 if x < 0, else H(x) = 1 for x >= 0.<br>$Ij = (\gamma_1+\gamma_2 j)H(\gamma_0-j)+(\gamma_3+\gamma_4 j+\gamma_5 j^2)H(j-\gamma_0)$ (12) |
| Voznak (2011) [53] | Based on E-model, $MOS(Ppl)'$ was proposed for the case of cascade codecs arrangement associated with several codecs (for example G.711 and G.729), as shown in (13), where $MOS(Ppl)'$ is the value after correction, $MOS(Ppl)$ is original MOS, $Ppl$ is loss rate, and a-e are constants.<br>$MOS(Ppl)' = MOS(Ppl)-\{a+[b \cdot (Ppl-c)^2-d]+(Ppl \cdot e)\}$ (13) |
| Goudarzi et al. (2011) [54] | $Ie$ functions for wideband SILK codec and narrow band SILK codec was proposed as shown in (14) and (15), where $Ie$ stands for the equipment impairment factor for SILK, and $B$ stands for a function of bitrate.<br>$Ie = (89983e^{-0.18B})/1000$ ; for Narrow band (14)<br>$Ie = 141-(11741B+458B^2-6B^3)/1000$ ; for Wideband (15) |
| Jiang & Huang (2011) [7] | The monitoring system was designed and implemented using the simplified E-model to evaluate VoIP quality, while $Ipacketloss$ and $Idelay$ shown (16) and (17), studied $Ipacketloss$ is the packet loss (%). The statistics received by the RTP packets is the total number of packets (N). The biggest serial number S and the smallest serial numbers can be extracted from serial number of sq field of RTP packet received. S2 is equal to S-s+1, the total number of data packets during the call. $Idelay$ is the average delay (ms), while A and B are the constants.<br>$Ipacketloss = (S2-N)/S2*100$ (16)<br>$Idelay = Ad+B$ (17) |
| Daengsi & Wuttidittachotti (2013) [43] | The Thai bias factor $and$ $MOS_{CQE-TH}$ was proposed for G.711 as shown in (18) and (19), where Thai bias factor is the factor obtained from subjective tests with a groups of native Thais, $MOS_{CQE}$ stands for MOS from an E-model tool, K is 5 for Thai, D stands for delay (s); 0s<delay<0.8s only, L stands for loss rate; 0 %<loss<10 % only, $MOS_{CQE-TH}$ stands for objective-MOS from E-model with Thai bias factor ($Bias_{TH}$).<br>$Bias_{TH} = -K(56.12-255.4L+88.48D-901DL-179.88D^2)/10^3$ (18)<br>$MOS_{CQE-TH} = MOS_{CQE} +Bias_{TH}$ (19) |
| Adel et al. (2013) [55] | Utilizing curve fitting technique, $MOS_C$ for or monitoring quality of multi-party VoIP communications was proposed as shown in (20), where $MOS_C$ and $MOS$ are the corrected E-model result and the stand E-model result in five-point scale, and $x_1$-$x_4$ are the parameters for G.711, ILBC or SLIK.<br>$MOS_C = x_1 MOS^3+x_2 MOS^2+x_3 MOS+x_4$ (20) |
| Assem et al. (2013) [4] | The improved the simplified E-model was proposed for several codecs (e.g. G.711 and G.726), as shown in (21) and (22), where $R$ is R-value, Ry is a second order function corrected utilizing curves fitted to PESQ scores, $Rx$ is a part of the simplified E-model which is corrected with PESQ points, and $a$-$c$ are codec constants for each codec.<br>$R = Ry-Id+A$ (21)<br>$Ry = a(Rx)2+bRx+c$ (22) |
| Giri & Tiwari (2015) [56] | The enhanced simplified E-model was presented with the proposed Id as shown in (23), where $Id$ is delay impairment factor, and $d$ is mouth to ear delay (ms).<br>$Id = -2.468 \times 10^{-14}d^6 + 5.062 \times 10^{-11}d^5 - 3.903 \times 10^{-8}d^4 + 1.344 \times 10^{-5}d^3 - 0.001802d^2 + 0.103d - 0.1698$ (23) |
| Triyason & Kanthamanon (2015) [57] | $Il$ associated with G.729 codec and packet loss was proposed for eight languages including Thai as shown in (24), where Il stands for language impairment factor, a-d are the constants for G.729 codec, and P stands for percent of random packet loss.<br>$Il = aP^3+bP^2+cP+d$ (24) |
| Triyason & Kanthamanon (2016) [58] | The proposed method used curve fitting to predict two parameters for the E-model, equipment impairment factor ($Ie$) and packet loss robustness factor ($Bpl$), for eight codecs (e.g. G.711, GSM, iLBC, Speex and Opus). |
| Wuttidittachotti & Daengsi (2016) [14] | The work proposed an $B_{TH-S}$ and $B_{TH}$ add-on to the E-model and simplified E-model respectively for G.711 referring to delay and loss, as shown in (25) and (26), where $B_{TH-S}$ and $B_{TH}$ are Thai bias factor for the simplified E-model and E-model with the effects of packet loss and delay, D stands for packet delay in second (s), while 0 s≤delay≤0.4 s, and L stands for packet loss rate ; 0≤loss≤0.1 (0.1 = 10 %).<br>$B_{TH-S}= -8.99+119.4L-35.78D-512.2L^2+125.6LD+379.6D^2+3240L^3+416.7L^2D-9.662LD^2-252.5D^3$ (25)<br>$B_{TH} = -6.674+224.8L-8.175D-3165L^2+50.02LD+28.49D^2+16920L^3+1112L^2D-151.8LD^2$ (26) |

| Wuttidittachotti & Daengsi (2016) [59] | Recently, a subjective MOS model and the bias factor for the simplified E-model enhancement for Skype was proposed, as shown in (27) and (28) respectively, where $MOS_{Subj}$ and $B_{TH}$ are subjective MOS and Thai bias factor, while $x$ stands for packet loss rate (%). |
|---|---|
| | $MOS_{Subj} = -(1/104)x^3 + (19/104)x^2 - (343/104)x + 4.3183$ (27) |
| | $B_{TH} = (13/104)x^3 - (1346/104)x^2 + 2.5141x - 0.5563$ (28) |

Unlike the subjective MOS prediction model in that previous study, the model in this study was developed from the raw data selected from only 9 scenarios under tests (four scenarios with delay of 800 ms were discarded due to the limitation of the simplified E-model which is that it cannot be used to predict VoIP quality if one-way delay is equal or more than 400 ms) [6, 48]. For the conversation-opinion tests, see Fig. 2, all tests were designed to be performed in two rooms with low background noise and low reverberation time, following ITU-T standards [34, 35], at the Central Library in King Mongkut's University of Technology North Bangkok. In this research, the 'real' VoIP system was applied for the conversation-opinion tests. The two IP phones used for testing were 'real' SIP phones. The VoIP system consists of Dummynet which is a network emulator and Asterisk which is a VoIP server. The subjective MOS from 250 KMUTNB students (134 male and 116 female participants, their average age was 21.83±4.59 years old) are shown in Table 2. Then, the validated data, 250 records total, were applied to create the new subjective MOS model for G.729 associated with loss (0-10 %) and delay (0-400 ms) effects utilizing the Surface fitting tool in MATLAB, which uses a regression technique. After consideration based on $R^2$ and RMSE, which stands for Root Mean Square Error, the equation as shown in (29) were chosen as the representative of the subjective MOS model, where $MOS_{G.729}$ is MOS voted by participants or subjects for G.729 codec, $x$ stands for loss rate (e.g. 0.03 for loss of 3 %) and $y$ stands for delay (e.g. 0.2 for delay of 200 ms). However, (29) is not exactly the same equation as previously presented in [13] because this study covered scenarios with and delay of 0 – 400 ms and loss of 0 - 10% only.

$$MOS_{G.729} = 4.113 - 13.960x - 0.2849y + 92.16x^2 + 18.030xy - 225.9x^3 - 165.7x^2y \quad (29)$$

V. SIMPLIFIED E-MODEL ENHANCEMENT

Since QoE is a subject of the end user's opinion, the simplified E-model that is an objective measurement method requires validation through the subjective results for its accuracy and reliability. Therefore, the subjective MOS prediction model has been used to enhance it in this research article. To use subjective MOS prediction model associated with G.729 codec for the simplified E-model enhancement, there are few tasks, as shown in Fig. 3, presented as follows:

Task#1: is to find subjective MOS values by using (8) to calculate the combination of loss of 0 %, 1 %, 2 %, 3 %, 4%, 5%,… , 10 % (where 10 % loss is two times the recommended acceptable loss rate following [4]) and delay of 0 ms, 50 ms, 100 ms, 150 ms,…, 400 ms (where 400 ms delay is the recommended acceptable rate of delay following [28]). Totally, there are 9 scenarios as in Table 2 to be calculated in this task for subjective MOS values before presenting in Fig. 4.

Task#2: there are two steps to be completed in this task, consisting of MOS-to-R conversion and R-value computation utilizing the simplified E-model. For MOS-to-R calculation, after prediction of subjective MOS in Task#1 using the data for Fig. 4, (3) has been used to convert from MOS values to R-values. Meanwhile the R-value computation utilizing the simplified E-model and its related equations, as in (5)-(7), with constants as proposed in [47] was applied, before combining with $I_{codec}$ (for, G.729, it is 10 [46]). Thus, those values change in a hundred-point scale.

Task#3: is subtraction of the R-values of the predicted subjective results from the R-values obtained from the simplified E-model before utilizing the different results to make the Bias function. It can be easily created by applying the regression technique, called the Surface fitting in MATLAB application, see Fig. 5. The Bias function could be in the form of polynomial equations. However, there are three candidate equations based on the Third-order polynomial with $R^2$ and RMSE values as shown in Table 3. The Bias function is shown as (30). Whereas $B_{TH}$ stands for Bias factor based on native Thai users, $a_1$-$a_9$ are the constants (see Table 4), $x$ stands for loss rate (for example, 3 for loss of 3 %), $y$ stands for delay (s) (for example, 200 for delay of 200 ms).

$$B_{TH} = a_1 + a_2x + a_3y + a_4x^2 + a_5xy + a_6y^2 + a_7x^2y + a_8xy^2 + a_9y^3 \quad (30)$$

Task#4: the last task, is about combining the result from the Bias function, as shown in (30) and the result from the simplified E-model, as shown in (5). That means, only the next time, this task will be applied to utilized to obtain the reliable result of the simplified E-model enhancement, defined as $MOS_{ES\text{-}CQE}$, which stands for the predicted MOS from the enhanced simplified – conversation quality estimation. Therefore, the improved model for G.729 can be used for VoIP quality prediction.

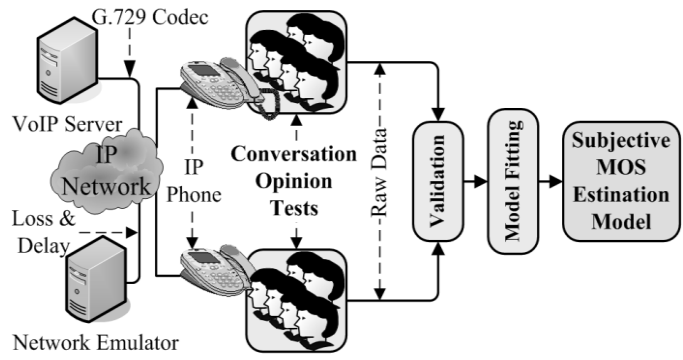

Fig. 2. Development of the Subjective MOS Prediction Model, adopted from [13]

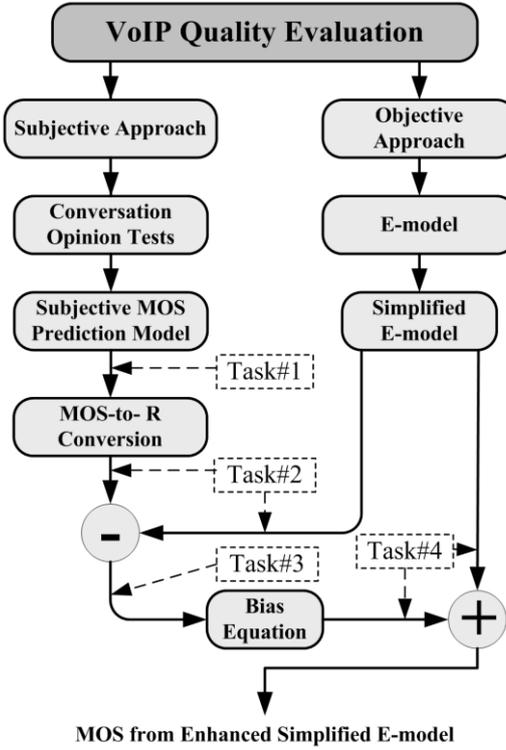

Fig. 3. Overview on simplified E-model enhancement, adopted from [2]

**Table 2**. Information of the selected scenarios under-test with G.729 from 250 participants, adopted from [13, 41]

| Scenario no. | Scenario | | No. of participants |
|---|---|---|---|
| | Loss (%) | Delay (ms) | |
| S01 | 0 | 0 | 24 |
| S02 | 2 | 0 | 30 |
| S03 | 4 | 0 | 24 |
| S04 | 6 | 0 | 30 |
| S05 | 10 | 0 | 24 |
| S06 | 0 | 400 | 28 |
| S07 | 3 | 400 | 32 |
| S08 | 5 | 400 | 30 |
| S09 | 10 | 400 | 28 |

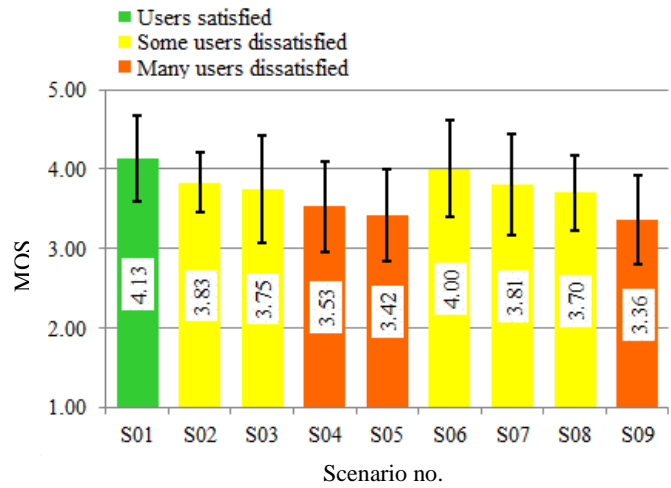

Fig. 4. Surface chart for Bias function of G.729 codec referring to loss (%) and delay (ms)

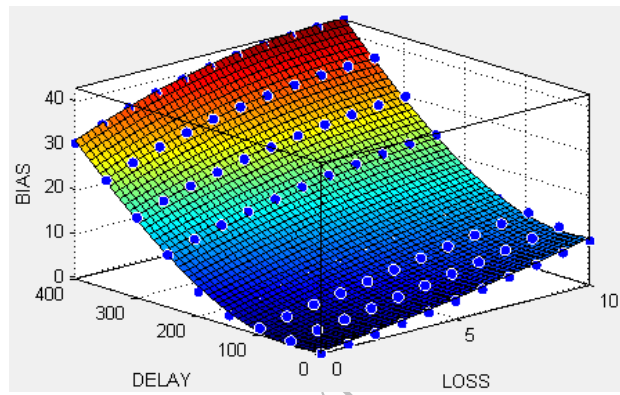

Fig. 5. Surface chart for Bias function of G.729 codec associated with loss (%) and delay (ms) effects with $R^2$ of 0.9964.

Table 3  The candidate equation based on 3$^{rd}$ order polynomial for Bias function representative

| Polynomial Order | | $R^2$ | RMSE | Remark |
|---|---|---|---|---|
| X | Y | | | |
| 3 | 2 | 0.9953 | 0.8872 | |
| 2 | 3 | 0.9964 | 0.7843 | *Selected* |
| 3 | 3 | 0.9964 | 0.7879 | |

Table 4  The constants for Bias equaion of G.729

| G.729 Codec Constant for Bias Function | |
|---|---|
| $a_1$ | 0.4327 |
| $a_2$ | 0.6654 |
| $a_3$ | -0.03461 |
| $a_4$ | 0.03563 |
| $a_5$ | 0.004689 |
| $a_6$ | 0.000379 |
| $a_7$ | -0.0004205 |
| $a_8$ | $-3.98/10^8$ |
| $a_9$ | $-2.52/10^7$ |

## VI. COMPARISON AND EVALUATION

In order to ensure the performance of the improved simplified E-model with the proposed idea, it requires comparison with the simplified E-model without enhancement. Nevertheless, with the assumption that subjective results are the most reliable and authentic, the closer to the subjective MOS prediction curve means the better model. Thus, the curves of the simplified E-model and the enhanced simplified E-model were compared with the curve of the subjective MOS prediction model referring to two parameters, loss of 0-12 % and delay of 0, 50, 100, 150, 200,… and 400 ms respectively) based on a hundred-point scale of R-value, as illustrated in Fig. 6 [2].

From the results as shown in Fig.6(a)-6(i), the R-value results from the enhanced model (the orange line with triangle marks) and the subjective MOS prediction model (the blue line with diamond marks) are almost the same in all scenarios under tests referring to the same loss and delay effects, whereas, the R-value results from the simplified E-model (the green line with square marks) are lower than the two models when the loss rates are increased, particularly the the simplified E-model results are dramatically lower than the two models when the delay rates are increased up to 300-400 ms. In detail, as shown in Fig. 6(a) referring to loss of 12 % and delay of 0 ms, the R-value results from the enhanced model and the subjective MOS prediction model are about 64-65 but the R-value result from the original simplified E-model is only 52 approximately, for example. When the delay rate is increased from 0 ms to 50, 100, 150, 200,… and 400 ms, it has been found that the gap between the simplified E-model and the two models is significantly increased more and more respectively. Especially, as illustrated in Fig. 6(i) referring to delay of 400 ms and loss of 12 %, the R-value results from the enhanced model and the subjective MOS prediction model are about 61-63 but the R-value result from the simplified E-model is only about 18.

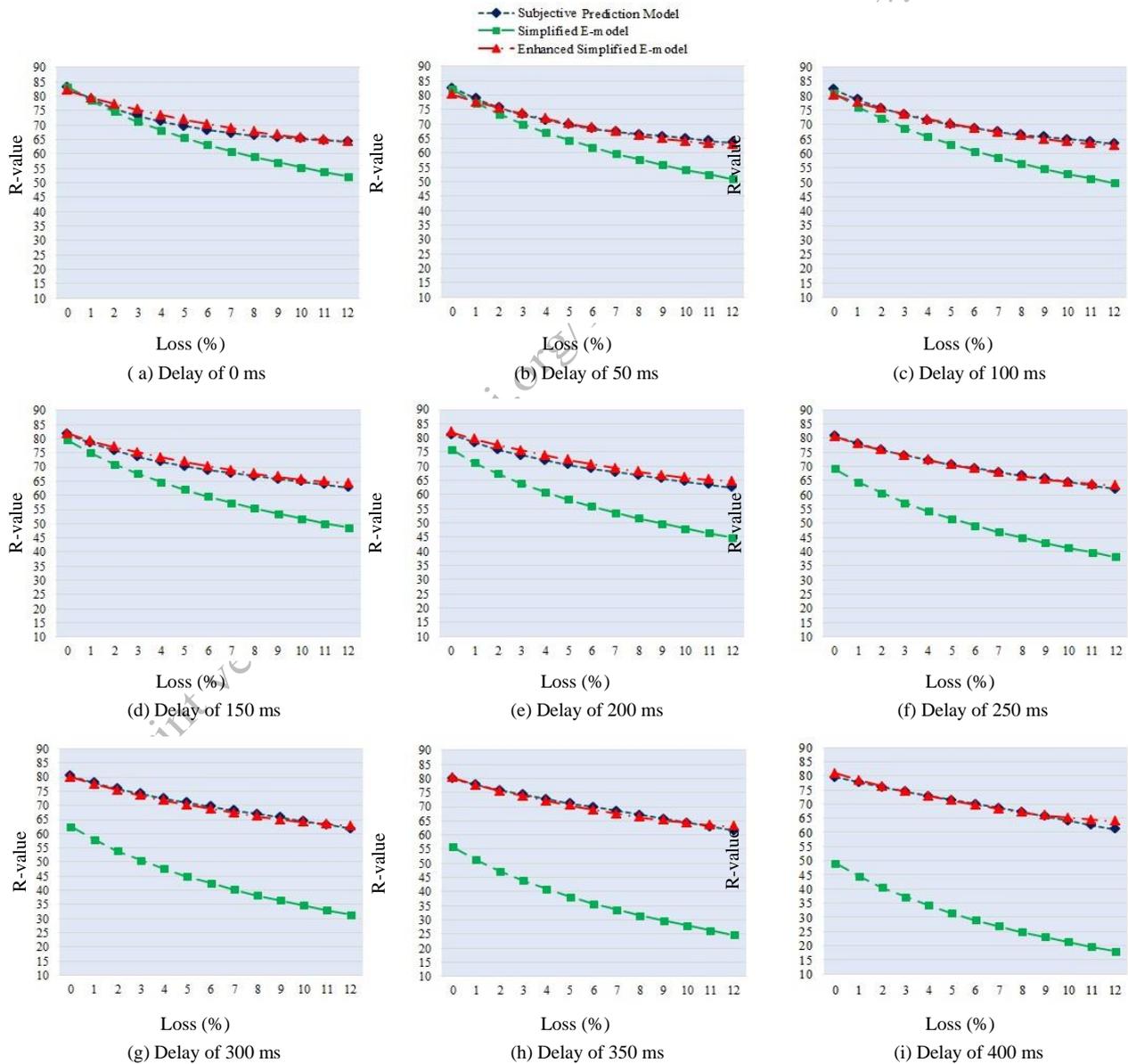

Fig. 6. Comparison of R-values from the enhanced simplified E-model with new coeffients, simplified E-model and the subjective prediction model for G.729 codec associated with 9 specific scenarios.

For more confidence with the simplified E-model enhancement, MAPE which stands for Mean Absolute Percentage Error equation, as shown in (31), and the test set (TS) were applied to this study [59]. Where MAPE stands for the mean absolute error in percent (%), $\bar{x}_i$ stands for the predicted value from the model, $n$ stands for the number of instances in the dataset and $x_i$ stands for the observed data or the subjective data in the meaning of this study.

$$MAPE = \frac{100}{n} \sum_{i=1}^{n} \left| \frac{\bar{x}_i - x_i}{x_i} \right| \tag{31}$$

The MAPE values from the simplified E-model and its enhanced model have been calculated using the special data referring to loss of 1 % and delay of 200 ms which was never published and the test set (TS) data applied from the same subjective MOS, as in [13], as mentioned in IV. Because of limitation about applying the data set from the previous work, the combination of the self-consistency test method and the split test method as mentioned in [60] have been applied, in order to separate the data into four test sets (TS1-TS4) for MAPE calculation. Each test set consists of one scenario related to Fig. 6(e), five scenarios and four scenarios related to Fig. 6(a) and 6(i) respectively. Of course, each set includes data of conversation-opinion test results obtained from both male and female participants who were 20-25 years old. The information about the test sets are shown in Table 5. For the scenario S1, S6 and S9, each scenario consists of data from 6 participants, while for the scenario S2-S5, S7-S8 and S10 consist of 7 participants per scenario. Totally, there are 67 records of data from 67 participants per test set. Then the data of the subjective MOS associated with Table 5 and the MOS results, called MOS-CQS, associated with the same scenarios obtained from the calculation using the original simplified E-model and the enhanced simplified E-model as shown in Table 6 were used to calculate MAPE using (31).

According to the meaning of MAPE values as described in [59, 61] that MAPE = 0 means a perfect forecast, 0 % < MAPE < 10 % means a highly accurate forecast, 10 % < MAPE < 20 % means a good forecast, and 20 % < MAPE < 50 % means a reasonable forecast, one can see in Table 7 that the MAPE values calculated from the simplified E-model results are 27.85 %, 28.32 %, 28.50 % and 29.30 % for TS1, TS4,TS3 and TS2 respectively, while the average MAPE is 28.47 %. Whereas, the MAPE values calculated from the results of the enhanced simplified one are 10.28 %, 11.96 %, 12.12 % and 12.49 % for TS2, TS3, TS1 and TS4 respectively. For the average MAPE, it is 28.47% obtained from the original simplified E-model, whereas it is only 11.71 % from the enhanced one. That means the original one is only a reasonable forecast, while the enhanced one is a good forecast with the error reduction of 58.87 %.

Thus, it can be claimed in this section that the enhanced simplified E-model utilizing the Bias function obtained from the subjective MOS prediction model is able to gain higher performance than the simplified E-model, although this enhanced E-model is only limited to delays of 0-400 ms and loss rates of 0-10 %.

## VII. DISCUSSION

This study proposed a model based on subjective MOS to gain higher performance of the simplified E-model, unlike previous works from other researchers that proposed their works based on objective methods, such as the following:

- several of them proposed the improved impairment factors, consisting of *Ie* and *Id* and the new jitter impairment factors [49, 52, 54, 56, 58]
- few of them proposed coefficients or constants of codecs for the simplified E-model [6, 47]

However, a few concepts to propose the language impairment factors as in [57, 58], which is equivalent to Bias function in this article, are similar but there were no subjective tests in their works. Excluding the original version of this work [2], the most similar works to this research are [14] and [59] which are the previous work before this research article which only focused on the G.711 codec and Skype which uses the SILK codec respectively. Therefore, this work fulfills the gap which has not been studied which is one of the most important ITU-T codec used in VoIP traffic, 93 % [18].

From the results shown in Fig. 6(a)-6(i), and as described in Section VI, it can be stated in this section that the original simplified E-model is very sensitive to delay effects because of the sensitiveness of the *Idelay* factor as in (6) which is an important part of the simplified E-model, while the propose model and the subjective MOS prediction model are not sensitive to those effects. Therefore, for an alternative solution, the *Idelay* factor might be investigated and improved in future work, instead of adding a Bias function to the simplified E-model as proposed in this study.

For its performance, after evaluating using graphs from Fig. 6 and the MAPE technique, one can see that the simplified E-model enhancement utilizing the Bias factor obtained from the different data from the simplified E-model and subjective MOS prediction model provided significant improvement. Particularly the simplified E-model for G.729 has been improved with error reduction of = 58.87 % (see Table 7). Thus, this VoIP-QoE model can be applied for VoIP quality prediction provided by G.729 with high confidence and reliability, then the predicted results can be applied for VoIP network and system management. This is the contribution of this work and it is the advancement beyond the previous version [2], although one may have questions about the limited scenarios (only S1-S10 with the test set TS1-TS4) of using MAPE for the evaluation. Moreover, not only providing the new constants (as in Table 4) of the proposed model as in (30) after validating and revising but also providing the MAPE technique for model evaluation, which has not been presented in [2], and the excellent performance of the proposed model when compared to the simplified E-model for G.729. Furthermore, the performance of this model with error reduction of more than 58

% is higher than the error reduction of more than 46 % provided by the enhanced simplified one for G.711 as in the previous research work [14], shown in Table 8.

Table 5  Information about the test set (TS) from conversation-opinion tests, adopted from [13]

| Scenario | Loss (%) | Delay (ms) | No. of Participants (per TS) | MOS-CQS±SD TS1 | TS2 | TS3 | TS4 | Remark |
|---|---|---|---|---|---|---|---|---|
| S1 | 0 | 0 | 6 | 4.00±0.63 | 4.17±0.41 | 3.86±0.38 | 4.29±0.76 | Related to Fig. 6 (a) |
| S2 | 0 | 400 | 7 | 4.00±0.82 | 3.86±0.38 | 3.86±0.38 | 4.29±0.76 | Related to Fig. 6 (i) |
| S3 | 1 | 200 | 7 | 4.14±0.38 | 4.00±0.58 | 3.86±0.69 | 4.00±0.58 | Related to Fig. 6 (e) |
| S4 | 2 | 0 | 7 | 3.71±0.49 | 3.86±0.38 | 3.86±0.38 | 3.86±0.38 | Related to Fig. 6 (a) |
| S5 | 3 | 400 | 7 | 3.88±0.64 | 4.25±0.71 | 3.50±0.53 | 3.63±0.52 | Related to Fig. 6 (i) |
| S6 | 4 | 0 | 6 | 4.17±0.75 | 3.67±0.82 | 3.67±0.52 | 3.50±0.57 | Related to Fig. 6 (a) |
| S7 | 5 | 400 | 7 | 3.57±0.53 | 3.71±0.49 | 3.71±0.49 | 3.71±0.49 | Related to Fig. 6 (i) |
| S8 | 6 | 0 | 7 | 3.43±0.53 | 4.00±0 | 3.57±0.79 | 3.29±0.49 | Related to Fig. 6 (a) |
| S9 | 10 | 0 | 6 | 3.17±0.41 | 3.83±0.41 | 3.33±0.52 | 3.33±0.82 | Related to Fig. 6 (a) |
| S10 | 10 | 400 | 7 | 3.43±0.53 | 3.43±0.53 | 3.43±0.79 | 3.14±0.38 | Related to Fig. 6 (i) |

Table 6  Simplified E-model and Enhanced Simplified E-model results based on the 10 evaluation scenarios

| Scenario | Loss (%) | Delay (ms) | Results Simplified E-model R-value | MOS | Enhanced Simplified E-model R-value | MOS |
|---|---|---|---|---|---|---|
| S1 | 0 | 0 | 83.200 | 4.139 | 83.633 | 4.149 |
| S2 | 0 | 400 | 49.103 | 2.528 | 80.191 | 4.033 |
| S3 | 1 | 200 | 71.265 | 3.656 | 79.471 | 4.004 |
| S4 | 2 | 0 | 74.646 | 3.807 | 76.552 | 3.889 |
| S5 | 3 | 400 | 37.160 | 1.927 | 74.659 | 3.807 |
| S6 | 4 | 0 | 68.270 | 3.515 | 71.934 | 3.690 |
| S7 | 5 | 400 | 31.503 | 1.672 | 71.950 | 3.685 |
| S8 | 6 | 0 | 63.186 | 3.263 | 68.894 | 3.548 |
| S9 | 10 | 0 | 55.336 | 2.856 | 65.986 | 3.403 |
| S10 | 10 | 400 | 21.238 | 1.290 | 64.417 | 3.326 |

Table 7  Evaluation results using the MAPE technique

| Test Set No. | Simplified E-model MAPE (%) | Average MAPE (%) | Enhanced Simplified E-model MAPE (%) | Average MAPE (%) | Remark |
|---|---|---|---|---|---|
| TS1 | 27.85 | 28.47% | 12.12 | 11.71% | Difference = (28.47-11.71)/28.47*100 = 58.87% |
| TS2 | 29.23 | | 10.28 | | |
| TS3 | 28.50 | | 11.96 | | |
| TS4 | 28.32 | | 12.49 | | |

Table 8  Comparison of model performance between the enhanced simplified E-model for G.729 and G.711

| Comparison | Codec | Average Error Reduction |
|---|---|---|
| Enhanced Simplified E-model versus Simplified E-model | G.729 | 58.87 % |
| | G.711 | 46.79 % |

## VIII. CONCLUSION

This research described processes for the simplified E-model enhancement for G.729 codec utilizing the subjective MOS prediction model based on native Thai users. This specific simplified E-model can be utilized within Thailand with high confident, accuracy and reliability. This approach has been ensured by performance evaluation using the MAPE technique with significant improvement, error reduction is about 58.9 %. This QoE-based work is an example for the enhancement of the VoIP quality measurement method, particularly for users in the countries that have their own tonal languages can conduct subjective tests with their native speakers by following the simplified E-model enhancement methods.

Nevertheless, this research work based on the simplified E-model should be extended to other codecs (for example G.726, and Opus) as future work. The impairment factors called *Ipacketloss* obtained from related works might be investigated in depth to see if it requires validation or not. Moreover, other impairment factors (e.g. jitter) should be studied using the same approach and applied to the enhanced simplified E-model in future work. Also, this QoE model should be implemented as a real-time tool or an application to predict voice quality of VoIP calls.

## ACKNOWLEDGMENT

Thank you to Rajamangala University of Technology Phra Nakhon and King Mongkut's University of Technology North Bangkok for supporting this research. Thank you to Ms. Nalakkhana Khitmoh and Mr. Jakkapong Polpong for the data set of subjective and objective results associated with G.729 codec. Thank you to Mr. Montri Rungruangthum for assistance with parts of the calculations. Lastly, thank you very much to Mr. Gary Sherriff for editing.